\documentclass[prl,twocolumn,balancelastpage,letterpaper,amsmath,floatfix]{revtex4} 

\usepackage{graphicx}
\begin{document}
\title{Time-Resolved Detection of Individual Electrons in a Quantum Dot}
\author{R. Schleser}
\author{E. Ruh}
\author{T. Ihn}
\author{K. Ensslin}
\affiliation{Solid State Physics Laboratory, ETH Z\"urich, 8093
Z\"urich, Switzerland}
\author{D. C. Driscoll}
\author{A. C. Gossard}
\affiliation{Materials Department, University of California, Santa
Barbara California 93106}
\date{\today}
\begin{abstract}
We present measurements on a
quantum dot and a nearby, capacitively coupled, quantum point contact
used as a charge detector. With the dot being weakly coupled to only
a single reservoir, the transfer of individual electrons onto and off
the dot can be observed in real time in the current signal from the
quantum point contact. From these time-dependent traces, the quantum
mechanical coupling between dot and reservoir can be extracted
quantitatively. A similar analysis allows the determination of the
occupation probability of the dot states.
\end{abstract}
\maketitle
\newcommand{\nf}{\normalfont}
\newcommand{\nm}{\,\text{nm}}
\newcommand{\muV}{\,\mu\text{V}}
\newcommand{\mV}{\,\text{mV}}
\newcommand{\eVV}{\,\text{eV}/\text{V}}
\newcommand{\Hz}{\,\text{Hz}}
\newcommand{\mK}{\,\text{mK}}

The electronic occupation in
semiconductor quantum dots can be read out using a quantum point 
contact (QPC) \cite{Field1993}.
Quantum dots are proposed as scalable spin qubits in a future quantum 
information
processor \cite{Loss1998} and the read-out could be implemented by a 
QPC detector.
Experiments using a radio-frequency single-electron transistor
resulted in a high bandwidth real-time read-out of a quantum dot's
charge state \cite{Wei-Lu2003}.
Theoretical considerations on dephasing \cite{Pilgram2002,Clerk2003}
have given evidence that a higher quantum measurement efficiency can
be obtained using a detector without
any internal degrees of freedom, such as, e.g., QPC containing a single mode.
Recent investigations using QPCs as charge detectors were performed on double
\cite{Elzerman2003PRB,Gardelis2003} and single
\cite{Hanson2003,Elzerman2004a} quantum dots, all using DC or
lock-in techniques that average over many electrons passing through
the dot. Very recently, real-time charge read-out measurements
on a split-gate defined structure were reported \cite{Elzerman2004}.

In this paper, we present measurements detecting single electrons in
real time using a QPC charge detector, in a circuit created entirely
by surface probe lithography \cite{Held1999, Luscher1999,Nemutudi2004}.
The structure consists of a quantum dot and a nearby,
electrostatically coupled QPC (see Fig. 1(a)).
It is written on a GaAs/Al$_{0.3}$Ga$_{0.7}$As heterostructure,
containing a 2-dimensional electron gas (2DEG) $34\nm$ below the
surface as well as a backgate $1400\nm$	 below the 2DEG, isolated
from it by a layer of LT-GaAs.
Negative voltages were applied to the surrounding gates (G1, G2,
S$_\text{QPC}$, D$_\text{QPC}$, the latter two also containing the
charge detection circuit), and to the back gate, to reduce the charge
on the dot and close its tunnel barriers. A voltage applied to gate P
was used to tune the detector QPC to a regime where it is sensitive to
the charge on the dot.

\begin{figure}[h!t!b!]
\includegraphics[width=3.3in]{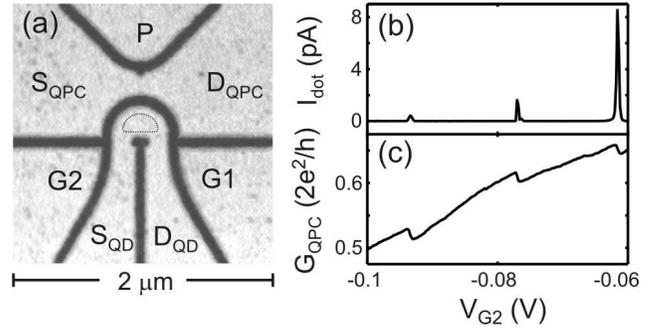}
\caption{(a) AFM micrograph of the structure with designations of
gates: source (S) and drain (D) of the quantum dot (QD) and the
quantum point contact (QPC) used as a charge detector; lateral gates
G1 and G2 to control the coupling of the dots to the reservoirs;
Plunger gate (P) to tune the QPC detector.
(b) Example measurement of the current through the dot.
(c) Simultaneous measurement of the conductance through the QPC,
where each step corresponds to a change of the dot's charge by one
electron.
\label{fig1}}
\vspace*{-0.6cm}
\end{figure}

All measurements were performed in a dilution refrigerator with a 
base temperature
of $80\mK$. 
A small bias voltage $V_\text{\textnormal bias, dot}=10\muV$ was
symmetrically applied across the dot between source (S$_\text{QD}$)
and drain (D$_\text{QD}$).
In a regime where both barriers are open and a transport current
through the dot is measurable, the example traces in Fig. 1(b) and 1(c)
were measured, showing the correlation between the Coulomb peaks in
the transport current (b) and the corresponding kinks in the
conductance vs. gate voltage curve of
the QPC (c). For later measurements, a compensation voltage 
$V_\text{P}=aV_\text{G1}+bV_\text{G2}$
was applied to the gate P, with constants $a$ and $b$ chosen
such as to keep
the charge detection circuit in a regime of almost constant
sensitivity.
  The sensitivity of our AFM-defined circuit is comparable
to similar set-ups realized by electron beam lithography defined
split-gate devices \cite{Elzerman2003PRB,Gardelis2003}.

The following measurements were performed with one tunnel barrier
(the one near the drain contact)
completely closed and the other one tuned to a very low electron
transition rate, of the order of only a few electrons per second.

Figure 2(a) shows a section of a measurement of the QPC detector's
conductance versus two different gate voltages, where each vertical
or horizontal trace can be thought of as being similar to Fig. 1(c).
Each step
corresponds to an electron being transferred onto the
dot. Towards lower values of $V_\text{G2}$ (see marked region), the smooth
behavior of the step transforms to a discrete appearance where only
two possible values for the QPC conductance are observed.

\begin{figure}[h!t!b!]		
\includegraphics[width=3.3in]{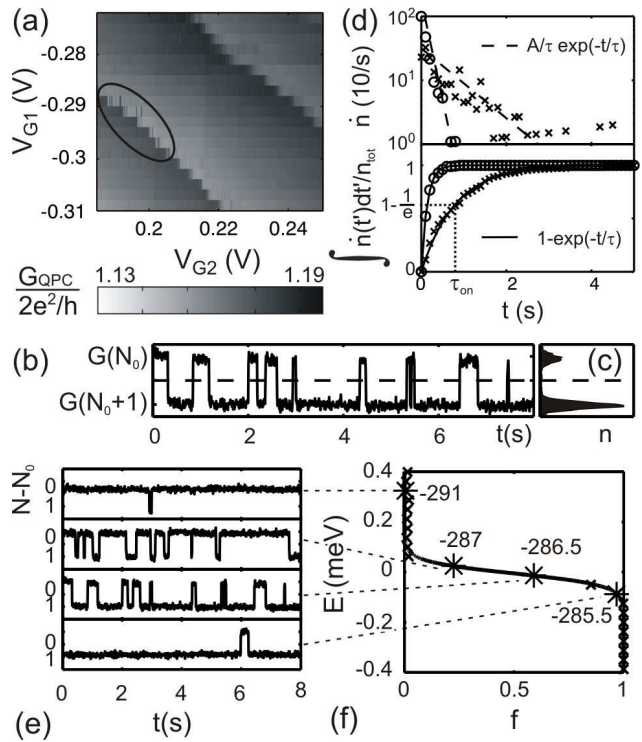}
\caption{
(a) Measurement of the QPC's conductance versus $V_\text{G1}$ and
$V_\text{G2}$. Towards lower values of $V_\text{G2}$ (see e.g.~inside the black
ellipse), the smooth, washed out appearance of the step in
conductance is replaced by a discrete switching behavior with only
two possible states. The time between individual measured points was
about $2$ seconds, the integration time of the (DC) measurement $0.2$
seconds.
(b) Single oscilloscope trace of the QPC's conductance
versus time.
The dot was tuned to a point near a step in the QPC's
conductance.
(c) Histogram of the data in plot (b) and 19 similar sweeps.
(d) The upper graph shows a histogram of the distribution of dwell
times of the electron both inside (o) and outside (x) the dot. A correction
was applied to account for the finite size of the oscilloscope's
measurement intervals. The bin size was chosen to be $0.1$ seconds,
data was taken from 20 sweeps of $9$ seconds each. The lower graph
shows the time-integrated, normalized distribution.
Dashed and solid lines represent exponential fits to the data with
the same parameters $\tau_\text{on}$, $\tau_\text{off}$ for both graphs.
(e) Changing the voltage $V_\text{G1}$
(at a constant $V_\text{G2}=185\mV$)	
changes the dot's
electrochemical potential and allows a transition from the $N$
electron state to the $N+1$ electron state. In the oscilloscope
traces, this is seen as a change in the relative occupancy of the two
possible QPC states.
(f) Distribution $f(E)$ extracted from
oscilloscope traces (where for every point, 20 traces each of length
9 seconds where taken into account). The data points marked by large
asterisks correspond directly to the traces shown in (b).
These asterisks
are labeled by the corresponding gate voltage $V_\text{G1}$ in mV. 
For a discussion of the energy scale see text.
\label{fig2}}
\vspace*{-0.3cm}
\end{figure}			

In the following, we present time-dependent traces of the QPC's
conductance. All traces were recorded using an oscilloscope at a
sampling frequency of $250\Hz$. Each trace had a length of 9 seconds.
For a time dependent measurement near a step in the QPC's
conductance, one observes that the system switches randomly between
the two states (Fig. 2(b)).
The difference between the two values corresponds to the observed
step height in Fig. 2(a) in the parameter range featuring a smooth
transition. This allows a discrimination between charging events on
the dot and other possible sources of switching events.

The upper plot in Fig. 2(d) shows an example for the distribution
of times the systems spends in both states.
The lower plot shows
the integrated, normalized distribution representing the probability
that after a certain time the system has changed its state at least once.
Exponential fits $\exp(-t/\tau)$ agree well with the data and suggest 
that (1) the behavior
of the system does not depend on its history and (2) that a single
energy level in the dot contributes to charging.

 From the exponential fits, two mean dwell times $\tau_\text{on}$ 
(electron on dot)
and $\tau_\text{off}$ (no electron on dot) can be obtained. The same numbers
can be extracted by counting the transitions per time interval
$f_\text{trans}$ during a time sweep and determining the
fractions $p_\text{on}$ and $p_\text{off}$ of the total
time the system spends in each of the two states (see histogram
in Fig. 2(c)). It follows that
$\tau_\text{on}=p_\text{on}/(f_\text{trans}/2)$ and
$\tau_\text{off}=p_\text{off}/(f_\text{trans}/2)$.

By changing the gate voltage $V_\text{G1}$, the electrochemical potential
in the dot can be modified. Figure 2(e) shows a series of time-dependent
traces for different $V_\text{G1}$. Using the lever arm
$\alpha_\text{G1}$, the gate voltage can be converted to an energy scale
$E=-\alpha_\text{G1}V_\text{G1}$, so that a distribution $f(E)\equiv 
p_\text{on}(E)$
is obtained (see Fig. 2(f)). To reduce the statistical
error, 20 time sweeps have been analyzed for every data point shown.

In a regime where only one tunnel barrier is open, no finite bias
transport measurements are possible which would allow the
determination of the exact value of $\alpha_\text{G1}$. Its value is therefore
obtained from the peak spacing by assuming a charging energy of the 
dot $E_\text{c}=2\mV$,
determined from finite bias transport measurements performed in a
more open regime of the dot. This method yields
$\alpha_\text{G1}\approx0.075\eVV$,
a value that was used in Fig. 2(f) and 3(c).	
Considering that, at lower electron numbers, $E_\text{c}$ tends to increase
due to the reduction in size of the dot, the energy and temperature
values to be determined below (see Figs.
2(f) and 3(c)) are to be considered an upper bound.

Motivated by the experimental findings and their statistical properties,
we interpret our results using the model depicted in Fig. 3(a), which
shows a single energy level in a quantum dot which can be aligned
with respect to the Fermi level of the reservoir, from which it is
separated by a tunable tunnel barrier.
The lower conductance in the QPC (see Fig. 2(b))
corresponds to the state where an
electron is on the dot, due to the capacitive coupling of the dot to
the QPC.
A similar scheme was used in an experiment using an SET as a
detector \cite{Wei-Lu2003}. A more sophisticated analysis considering
more than one level in the dot can be performed using the theory by
Beenakker \cite{Beenakker1991}.

\begin{figure}[h!t!b!]
\includegraphics[width=3.3in]{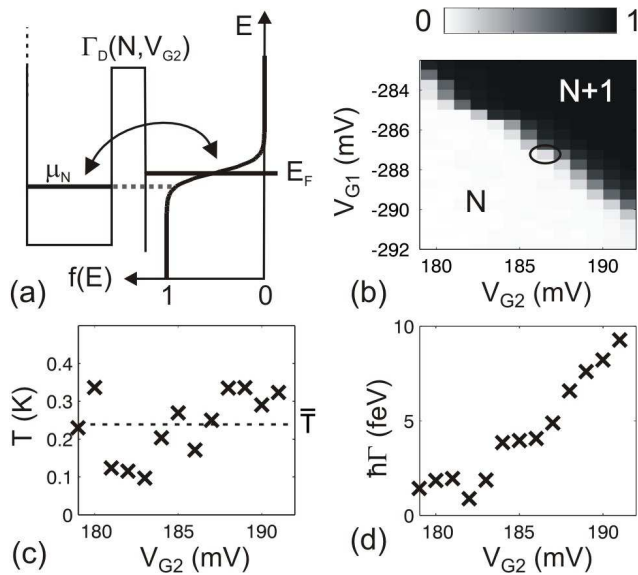}
\caption{
(a) Model for the transfer of individual electrons between a
single revervoir and a quantum dot. The tunnel coupling is assumed to
depend on the single-level state $N$ and can be tuned via the voltage
$V_\text{G2}$.
(b) Greyscale plot of the relative occupancy
$p_\text{on}$ of the
dot. Each point is calculated from 20 oscilloscope traces, each $9$
seconds long, via the method described in the text. The section
corresponds roughly to the marked range of Fig. 2(a). Due to a minor charge
rearrangement, however, the voltage ranges are not exactly the same.
The ellipse marks the point where the statistical analysis of dwell
times (Fig. 2(d)) was applied.
This measurement was performed with the QPC containing more than one mode.
(c) Temperature extracted from the data in (a) using a fit of the
Fermi function (see Fig. 3(c)) for every value of the gate voltage
$V_\text{G2}$.
(d) Coupling strength $\Gamma$ to the leads extracted from the same
set of data as (a) and (b). As expected, $\Gamma$ increases with
$V_\text{G2}$.
\label{fig3}}
\vspace*{-0.5cm}
\end{figure}

Assuming that an electron is on the dot, the mean rate at which it
will leave the dot is
$
    \tau^{-1}_\text{off}=\Gamma\times\left(1-f(\mu_N)\right)\;,
   \label{eqtauoff}
$
where $\tau_\text{off}$ is the average time interval
the dot stays in the $(N+1)$ electron state,
$\mu_N$ is the electrochemical potential for the addition of the Nth electron,
$f(E)$ is the distribution function in the reservoir,
and $\Gamma$ is the coupling between the
level and the reservoir. Physically, the value of $\Gamma$ accounts for
the strength of
tunnel barriers, wave function overlap and the lead's density of states
(assumed to be constant over the relevant energy interval).
Correspondingly, the rate for electrons to
tunnel on an (initially empty) dot is
$
	\tau^{-1}_\text{on}=\Gamma \times f(\mu_N).
   \label{eqtauon}
$
It follows 
that
$
\Gamma=\tau_\text{off}^{-1}+\tau_\text{on}^{-1}
\label{eqGamma}
\text{and}\;\;\; f(\mu_N)=\tau_\text{off}/(\tau_\text{off}+\tau_\text{on})\,.
$

This means that we are able to determine the tunnel coupling
$\Gamma_N$ of an individual energy level to a single reservoir as
well as the energy distribution $f(E)$ of the lead.

In Fig. 2(f), the
solid line represents a fit using the Fermi distribution
$f(E)=1/(1+\exp(-E/k_\text{B}T))$, from which a temperature $T\approx 
150\mK$ can be
extracted.

In the following we
present data in a more extended parameter regime.
Figure 3(b) shows a plot of the extracted distribution function 
$f(E)$ versus the two gate voltages $V_\text{G1}$ and $V_\text{G2}$. For each
scan in the direction of $V_\text{G1}$, a fit using the Fermi distribution
was made. The resulting temperature values are represented in Fig.
3(c), the mean value being slightly above $200\mK$. The origin of the
scattering of $T$ values in Fig. 3(c) remains to be investigated,
since it can not be
assigned entirely to uncertainties in the fitting procedure or lack
of data points.

The numerical extraction of the parameter $\Gamma$ is most reliable
near the transition.
$\Gamma$ was therefore determined where $f(E)$ was
closest to $1/2$. The resulting curve is shown in Fig. 3(d). Over the
range presented, $\Gamma$ changes by roughly one order of magnitude.
Further reducing $V_\text{G2}$ by only a few ten mV leads to a 
further increase of
the observed times between switching events (of the order of minutes 
and more). 
This is in accordance with recent observations of long dwell times of
electrons on a single \cite{Cooper2000} or double \cite{Gardelis2003}
dot.

We have reported on the real-time charge read-out of a
quantum dot using a quantum point contact, both defined by AFM Lithography.
We estimate that amplifier and cabling bandwidth as well as the sensitivity
can be improved to reach read-out frequencies in the MHz range
(see also \cite{Elzerman2004}).


\end{document}